\documentclass[modern, times, trackchanges]{aastex61}

\usepackage{booktabs, gensymb}
\newcommand{\NA}{---}
\accepted{11 March 2020}
\submitjournal{ApJ}

\shorttitle{Necroplanetology}
\shortauthors{Duvvuri, Redfield, \& Veras}

\begin{document}

\title{Necroplanetology: Simulating the Tidal Disruption of Differentiated Planetary Material Orbiting WD 1145+017}

\correspondingauthor{Girish M. Duvvuri}
\email{girish.duvvuri@gmail.com}

\author[0000-0002-7119-2543]{Girish M. Duvvuri}
\affiliation{Department of Astrophysical and Planetary Sciences, University of Colorado, Boulder, CO 80309, USA}
\affiliation{Department of Astronomy and Van Vleck Observatory, Wesleyan University, Middletown, CT 06459, USA}

\author[0000-0003-3786-3486]{Seth Redfield}
\affiliation{Department of Astronomy and Van Vleck Observatory, Wesleyan University, Middletown, CT 06459, USA}

\author[0000-0001-8014-6162]{Dimitri Veras}
\altaffiliation{STFC Ernest Rutherford Fellow}
\affiliation{Department of Physics, University of Warwick, Coventry CV4 7AL, UK 5}
\affiliation{Centre for Exoplanets and Habitability, University of Warwick, Coventry CV4 7AL, UK}



\begin{abstract}
The WD 1145+017 system shows irregular transit features that are consistent with the tidal disruption of differentiated asteroids with bulk densities $< 4\, \rm g \, \rm cm^{-3}$ and bulk masses $\lesssim 10^{21} \, \rm kg$ \citep{Veras2017}. We use the open-source N-body code \texttt{REBOUND} \citep{REBOUND} to simulate this disruption with different internal structures: varying the core volume fraction, mantle/core density ratio, and the presence/absence of a thin low-density crust. We allow the rubble pile to partially disrupt and capture lightcurves at a specific point during the disruption at cadences comparable to those from ground-based photometry. As a proof-of-concept we show that varying these structural parameters have observationally distinguishable effects on the transit light curve as the asteroid is disrupted and compare the simulation-generated lightcurves to data from \cite{Gary2017}. With the caveat that our simulations do not model the sublimation in detail or account for its effects on orbital evolution, we find that a low core fraction and low mantle/core density ratio asteroid is most consistent with the stable transit feature present for multiple weeks circa April 2016 (referred to as G6121 in \citet{Gary2017} and A1 in \citet{Hallakoun2017}). Connecting tidal disruption simulations to photometry suggests characteristics for the interior structure and composition of an exoplanetary body, information that is only possible because we are observing the death of the planetary system in action. All-sky survey missions such as \emph{TESS} and \emph{LSST} will be able to detect other systems like WD 1145+017, creating a sample of subjects for a new subfield of planetary science: necroplanetology.
\end{abstract}




\section{Introduction} \label{sec:intro}
The first exoplanet discovery showed that planets can be found orbiting post-main-sequence objects like pulsars \citep{WolszczanFrail1992} and the \emph{Kepler} mission showed that most main-sequence stars have at least one planet \citep{Cassan2012, KeplerOccurrence}. Combined with the observed infrared excess, detectable dusty disks, and metal-polluted spectra of $\sim4\%$, $\sim2\%$, and $\sim30\%$ of white dwarfs respectively \citep{Zuckerman2003, Zuckerman2010, Koester2014, Farihi2016review}, we know that planetary systems survive stellar evolution in some form for a significant fraction of stars. The ``polluted" spectra refer to metal lines which must be caused by recently accreted material since the high surface gravity of a white dwarf would sink metals beneath the photosphere over the course of weeks \citep{Schatzman1948}. There are discrepancies in the observed chemical composition, spatial and kinematic distributions, and line strengths  of the polluted material compared to what would be expected for accretion from the interstellar medium, identifying the presumed source of the pollution as planetary debris orbiting the white dwarf instead \citep{Jura2003, KilicRedfield2007, FarihiRedfieldISM}.

Many previous studies have used high-resolution spectroscopy of these polluted white dwarfs to reconstruct the chemistry of these former exoplanetary objects \citep[e.g.,][]{chemistry_1, 2012MNRAS.424..333G, 2014AREPS..42...45J, chemistry_2, chemistry_3, 2018MNRAS.479.3814H, 2018MNRAS.477...93H, Doyle2019, Bonsor2020}, obtaining information about their bulk composition at a level unmatched by current exoplanetary characterization techniques applied to main-sequence stars. But these chemical analyses still arrive late to the scene: the planetary material has been destroyed and accreted already so that the characteristics of individual bodies cannot be determined \citep{Veras2016}. During Campaign 1 of the \emph{K2} mission \citet{Vanderburg2015} reported the discovery of irregular transits caused by what appeared to be an asteroid or minor planet still being disrupted by its host WD 1145+017. Follow-up photometry from the ground showed that there were multiple periodic signals of varying mean anomaly, shape, asymmetry, and depth clustered around a $\sim 4.5$ hour orbit \citep[e.g.,][]{Vanderburg2015, 2016ApJ...818L...7G, Rappaport2017, Gary2017}.

\citet{Xu2016} found circumstellar absorption in the polluted metal lines which \citet{Redfield2017} observed to vary on both minute timescales corresponding to transits coincident with the spectroscopic observations and over the course of months due to a mechanism that remains uncertain. \citet{Redfield2017} proposed a warped eccentric disk interior to the transiting material that is consistent with most of the observed features while \citet{Farihi2017} proposed a magnetospheric accretion process. However, the \citet{Farihi2017} model requires magnetic fields higher than the limits imposed by the X-ray flux emanating from WD 1145+017 \citep{Farihi2018}. \citet{Cauley2018} amends the \citet{Redfield2017} model by describing a series of 14 individual narrow disks that share the same focus but have different eccentricities and apsidal angles that increase linearly with distance from the white dwarf. \citet{Vanderburg2018}, \citet{Xu2019}, and \citet{Fortin2020} discuss further theoretical and observational work on the gas disk and its relationship with the transiting debris.

\citet{Gurri2017} used N-body simulations to place limits on the eccentricity and mass of the transiting objects under certain assumptions of mass ratios and restricted drift in the orbital period. Complementing this study, \citet{Veras2017} used the N-body code \texttt{PKDGRAV} \citep{2000Icar..143...45R, pkdgrav} to simulate the disruption of an individual asteroid to place constraints on the bulk density and show that the transits were best explained by a differentiated body orbiting within the Roche limit for its mantle density but outside the limit for the core. However, these simulations only tested one possible differentiated internal structure to compare to a homogeneous rubble pile. This paper employs a similar methodology but expands it to explore a variety of internal structures and generate lightcurves that can be compared to data. In Section \ref{sec:sim_params} we describe the parameters for this suite of simulations and in Section \ref{sec:lightcurves} we present the lightcurves generated to show that structural differences lead to observationally distinguishable features in the lightcurve. Section \ref{sec:fitting} reports our best fit for a specific transit lightcurve from \citet{Gary2017}. We conclude in Section \ref{sec:conclusion} by discussing the likely structure of the disrupting material transiting WD 1145+017, relating our work to the \citet{Cauley2018} model, and the implications of using this method to study the interior structure of other exoplanetary bodies.

\section{Simulation Setup}\label{sec:sim_params}
\begin{figure}[ht!]
\plotone{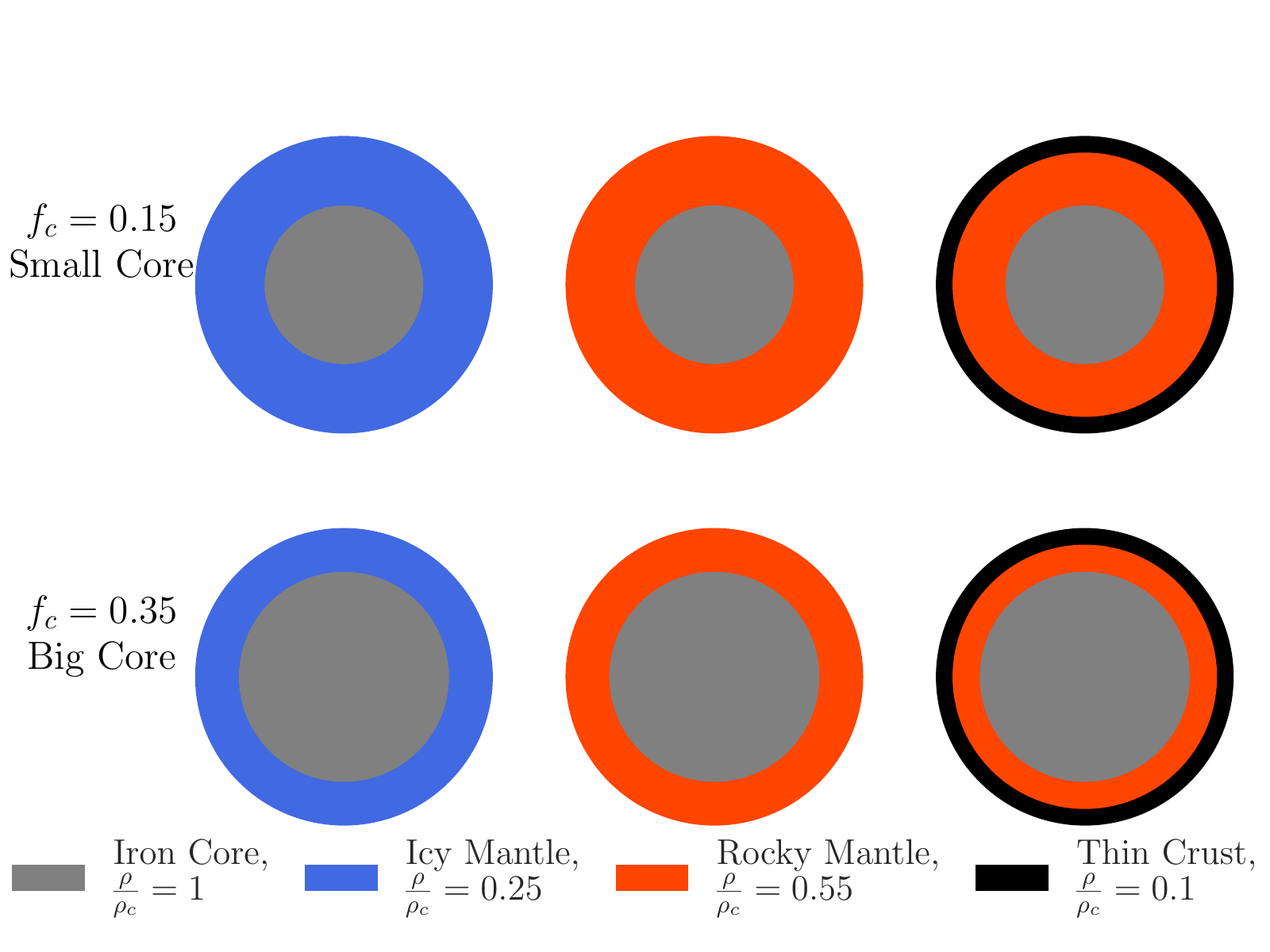}
\caption{These are six examples of asteroids that we use to demonstrate the structural differences we explore in our rubble pile simulations. We can change the fractional volume of the core $f_c$, the density of the mantle relative to the core $\frac{\rho_m}{\rho_c}$, and decide whether or not to include a thin low-density crust. This cartoon shows the boundary values for both $f_c$ and $\frac{\rho_m}{\rho_c}$, but there are additional structures at an intermediate value for both parameters. For all structures, the bulk density of the rubble pile is either $3 \ \rm or \ 4 \, \rm g\, \rm cm^{-3}$. All possible permutations lead to a total number of 36 rubble pile simulations.}
\label{fig:structure_cartoon}
\end{figure}

We set up the disruption simulations by randomly packing $N=3000$ spheres of equal radius, creating a single rubble pile with $M_b = 10^{24}\, \rm kg$ and no spin. This higher mass is chosen for computational considerations discussed later in the text. The rubble pile structure is allowed to vary between all permutations of core volume fraction $f_c = 0.15, 0.25, \textrm{ and } 0.35$; mantle/core density ratio $\rho_m / \rho_c = 0.25, 0.40 , \textrm{ and } 0.55$; the presence or absence of a thick crust with volume fraction $f_l = 0.1$ and crust/core density ratio of $\rho_l / \rho_c = 0.1$; and bulk density $\rho = 3 \textrm{ or } 4 \, \rm g \, \rm cm^{-3}$. By fixing both the bulk mass and density, the bulk radius is determined and the radii of the individual particles are scaled appropriately. For each simulation, the rubble pile was placed in a circular $e=0$, edge-on $i=90\degree$ orbit with a semimajor axis $a=0.0054 \, \rm AU$ around a white dwarf with $M_\star = 0.6 \, M_\odot$ to be consistent with \citet{Veras2017}. At the time the simulations were performed, this was only a fiducial mass estimate for WD 1145+017, but \citet{Izquierdo2018} report the mass of WD 1145+017 as $0.63 \pm 0.03 M_\odot$. The parameters common to all of the final simulations are listed in Table \ref{table:common_params} while Table \ref{table:sim_param_table} lists the parameters that were varied between simulations. Figure \ref{fig:structure_cartoon} is a cartoon illustration demonstrating how changing each of these parameters affects the structure of the simulated rubble pile.

Previous studies of tidally disrupted differentiated bodies \citep{Leinhardt2012, Veras2017} used the same structure consistently: $f_c = 0.35$, $\rho_m / \rho_c = 0.25$ and no crust. However, \citet{Carter2015} showed that when simulating planetesimal growth via core accretion, the final core fractions spanned the complete range between $0$ and $1$ regardless of orbital period with most planetesimals ending up with a core fraction between $0$ and $0.4$, motivating the range of $f_c$ values chosen in this study. They also showed that the relative mass fractions of the core and mantle varied widely but strongly correlated with orbital period. Since we cannot be certain where the disrupting bodies originally formed, we adopt the three values above to roughly represent a mantle that lies between an ``icy" volatile-rich and a ``rocky" Earth-like density with a ``metallic" iron-rich core. These names are only suggestive and a single parameter for the mantle/core density ratio cannot capture the details of the possible compositions and substructures of the differentiated layers. The two choices of bulk density were motivated by the limits placed by \citet{Veras2017} when previously modelling the WD 1145+017 system. We include a crust to see whether or not multiple layers of differentiation could be detected.

Using the leapfrog integrator of the $N$-body simulation code \texttt{REBOUND} \citep{REBOUND}, we integrate these 36 simulations for 100 orbits with $P = 4.5$ hours, $e=0.0$, $M_\star=0.6 \, M_\odot$, and an integration timestep of $dt=10$ seconds. We record a snapshot of the simulation after the second orbit using the Simulation Archive feature of \texttt{REBOUND} \citep{SimulationArchive} then continue to evolve this snapshot for a half-orbit while saving the positions of each particle for each timestep. By projecting the particles against the face of the white dwarf divided into $\sim 50000$ pixels and counting the pixels obscured by a simulation particle, we generate light curves for each simulation at a cadence comparable to the ground-based photometry available for the system. We smooth the scatter in the pixel-counted lightcurves using a Savitzky-Golay filter \citep{SavitzkyGolay} before fitting to the photometry. We fit these template lightcurves to the photometry by scaling the depth and duration of transit with two free parameters. Scaling the depth is equivalent to inflating all particles equally, differing from inflating only the particles separated from the bulk of the asteroid as \citet{Veras2017} did.

While the bulk density determines whether or not tidal disruption occurs, \citet{Veras2014} showed that the timescale of disruption $t_{\text{fill}}$, the time it takes for a rubble pile to completely disrupt and fill out a debris ring, depends on the bulk radius $R_b$. For our chosen $\rho_b$ and $M_b$, the snapshot light curve corresponds to a $t/t_{\text{fill}} \approx 0.06$. Theoretically, these light curves should match any identically structured rubble pile of a different mass at roughly the same $t/t_{\text{fill}}$. We verified this for 5 values of $M_b \in [10^{18}, 10^{24}] \, \rm kg$ and 10 random seeds for Structure 1. This verification was not done for every simulation because a smaller $M_b$ results in smaller particles, and the face of the white dwarf must be divided into more pixels to count these particles accurately, slowing down the calculation of the light curve. Testing a range of masses for each of the 36 structures was not computationally tractable. The random seed tests also allowed us to ensure that the observed lightcurve differences were genuinely caused by the structural differences in the rubble pile and not simply by the particular random packing. Unfortunately, the high mass we chose for computational purposes sped up the the disintegration of the rubble pile such that only two orbits after the \texttt{REBOUND} snapshot had meaningful transit signatures before creating a debris ring. We only used the first transit for fitting to lightcurves as described in the following section, severely limiting our ability to study the temporal evolution of the transit feature.

\section{Disruption Lightcurves}\label{sec:lightcurves}

\begin{figure}[htp]
\epsscale{1.1}
\plotone{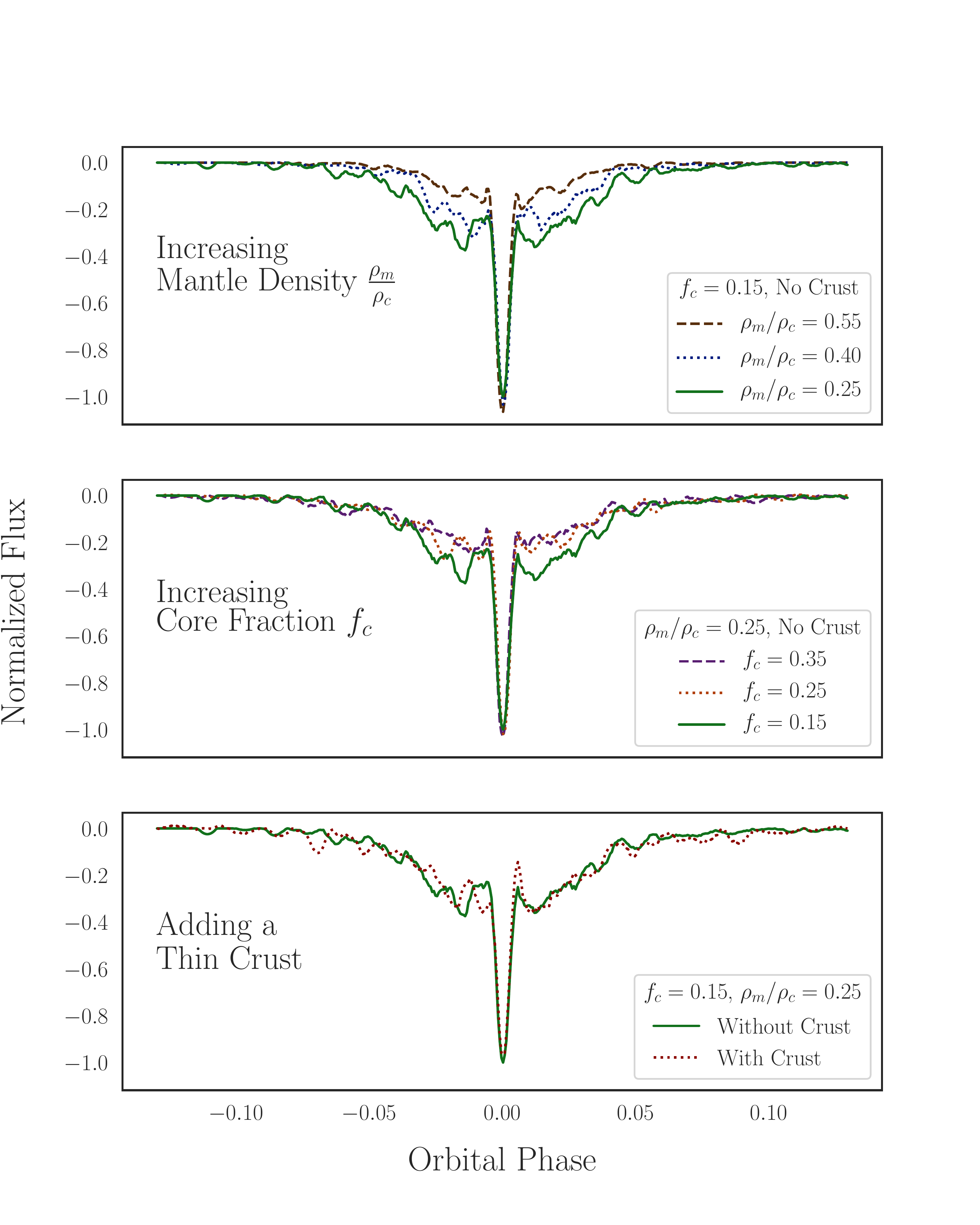}
\caption{Specific structural changes do lead to distinguishable differences in the lightcurve, but require high signal-to-noise ratio (SNR) observational data to confidently rule out regions of parameter space. Both the core fraction $f_c$ and the mantle/core density ratio $\frac{\rho_m}{\rho_c}$ control the strength of the wings, but the core fraction $f_c$ has an additional effect on the asymmetry of the transit light curve. The presence of a crust adds some scatter to the lightcurve, but this effect is marginal and highly unlikely to be constrained by observations.}
\label{fig:comparison}
\end{figure}

\citet{Canup2010}, \citet{Leinhardt2012}, and \citet{Veras2017}  have previously described the geometry of mantle disruption. First the entire rubble pile is stretched into a lemon shape as tidal forces act more strongly on the rarer mantle material. When the self-gravity of the mantle material is no longer sufficient to maintain cohesion, particles stream out from the end closer to the gravitational host at the L1 point and then at the farther end at the L2 point. This streaming is initially intermittent and slow, then speeds up as the number of gravitationally-bound mantle particles decreases, pushing some particles into a slightly shorter orbital period and others into a slightly longer one. If the core is safely beyond its Roche limit, it remains safely intact, but otherwise breaks up and spreads out to fill out the ring. When \citet{Veras2017} simulated this particular system, they found that there was a difference in the rate of material loss between the two streams and eventually the space between them filled out to form one ring. Our simulations are consistent with \citet{Veras2017} in this regard. The time at which we capture the lightcurve corresponds to when some of the mantle has already been stripped but before it has completely dispersed into a ring.

Figure \ref{fig:comparison} shows how changing individual structural features changes the lightcurve. Increasing the mantle/core density ratio weakens the wings relative to the core since less of the mantle has been removed from the parent body. Increasing the core fraction has a similar effect but also decreases the asymmetry of the lightcurve caused by the L1/L2 mass loss difference. This trend could be caused by either the increased gravitational potential from the core or because there is less mantle material to lose and the difference becomes less observably significant. Adding a crust has a minimal effect by only adding a little more scatter to the wings, such that the presence or absence of a crust is least constrained by fitting to the photometry.

\section{Fitting}\label{sec:fitting}
\begin{figure}[htp]
\epsscale{1.1}
\plotone{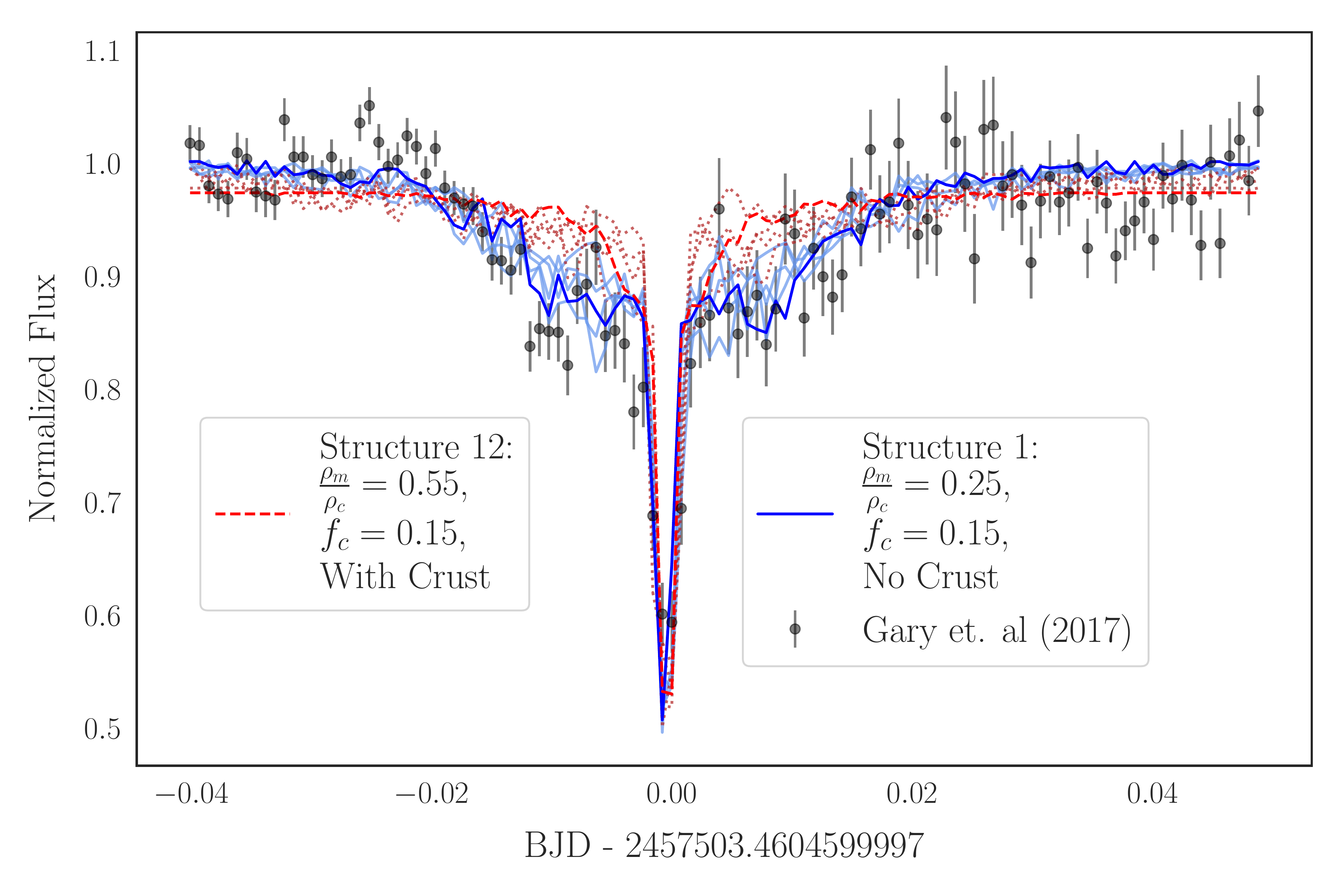}
\caption{Structure 1 (which has a bulk density $\rho_b = 3 \, \rm g \, \rm cm^{-3}$ mantle/core density ratio $\rho_m / \rho_c = 0.25$, core fraction $f_c = 0.15$, and no crust) has the best fit among all structures and is plotted as a solid blue line. The four next best fitting structures, bold-faced in Table \ref{table:sim_param_table}, all have $f_c=0.15$ and are plotted in pale blue solid lines. Structure 12, which has the same bulk density and core fraction but a higher mantle/core density ratio $\rho_m / \rho_c = 0.55$, is the worst fit among all the structures and is plotted as a dashed red line. Structure 1 reproduces the key features of this transit: a sharp central trough, asymmetric wings, and scatter in the extended ingress and egress. Structure 12 reproduces the sharp central trough but the high density of the mantle prevents strong asymmetric wings. The four next worst fits, italicized in Table \ref{table:sim_param_table}, are plotted in pale red dotted lines. Like Structure 12, they are all much more symmetrical and have weaker wings than the data.}\label{fig:gary_fit}
\end{figure}

\citet{Gary2017} presented lightcurves of WD 1145+017 from 105 observing sessions over six months starting on 2016 January 25. These observations were taken from multiple locations and with different telescope setups. We found a recurring feature in the April observations taken with the IAC80 telescope on the Canary Islands at a cadence of $\sim 68$ seconds. This feature strongly resembled our simulated lightcurves with a single well-defined dip flanked by asymmetric wings. Other features exist that appear to be blended superpositions of dips or do not persist with a coherent shape over multiple orbits. Our simulations only account for the tidal disruption of a single rubble pile and do not consider detailed physics in the sublimation of particles making them ill-suited to describing these more complex transit features \citep{Veras2015, Xu2018, Xu2019}. This particular transit may have been early enough in the disruption of the transiting body or symmetrical enough in its sublimation such that its orbital evolution did not diverge significantly from purely tidal disruption. Combined with its isolation from other features, this transit observation was ideal to demonstrate our proof-of-concept comparing differentiated disruption simulations to photometry.

We use smoothed lightcurves from each structure to fit to the 26\textsuperscript{th} April 2016 transit of this feature in the data from \citet{Gary2017} and crudely approximate the sublimation of the particles by scaling the transit depth by one free parameter that describes an identical effective obscuration area for all particles. Since we cannot identify exactly when this particular feature's progenitor began disrupting, we similarly scale the transit duration by another free parameter which assumes that the transit shape is preserved as the ring fills out. Therefore our simulated lightcurves can only describe the early behavior of the transiting body, before sublimation affects the orbital evolution of individual particles significantly. All fits were performed with the affine-invariant MCMC algorithm proposed by \citet{GoodmanWeare2010} and implemented in Python by the \texttt{emcee} package described in \citet{emcee}. The four free parameters were the time at which the transit starts, the factor by which to scale the total duration of the transit, the amplitude by which to scale the depth of the transit, and the out of transit baseline flux.

Figure \ref{fig:gary_fit} shows the best-fitting structure which had a bulk density of 3 g cm$^{-3}$, a core fraction $f_c = 0.15$ and a mantle/core density ratio $\rho_m / \rho_c = 0.25$. The five best fitting structures all shared a small core fraction $f_c = 0.15$. The poorest fits have higher core fractions and/or a higher mantle/core density ratio because these parameters cannot reproduce the asymmetry and strong wings that were observed in this transit. The chi-squared statistic for the best-fit from each structure is listed in Table \ref{table:sim_param_table}. We note that these fits cannot be used to infer structural parameters with much confidence because we do not model the relative sublimation of material from different layers or account for composition or orbital evolution due to sublimation. There are some trends in better fits correlating to lower core fraction, but the different mantle densities of the best fits show that we cannot uniquely specify the structure of an asteroid using our simulations. Nevertheless, these simulations show that over similar dynamical timescales, the internal structure of a disrupting rubble pile asteroid will impact the spatial distribution of particles from these layers which should have observable consequences for the transit light curve.

\section{Conclusion}\label{sec:conclusion}
The WD1145+017 system offers (currently) unique insight into the fate of post-main-sequence planetary systems. The options for post-main-sequence planetary evolution seem largely limited to: feeding circumstellar disks; accreting onto the host white dwarf; or being ejected into the interstellar medium: a possible origin for the recently discovered interstellar visitor `Oumuamua \citep{Oumuamua1, Oumuamua2, Rafikov2018}. Pulsar planets and WD 1145+017 are the rare exceptions discovered so far although other stable post-main-sequence planetary configurations could exist \citep{Veras2016}.

Our modelling of the photometry from this singular system suggests that the bodies creating the observed transit signatures are differentiated with a very small core fraction and low density mantle, resembling an asteroid with a partially differentiated structure and volatile-rich mantle like Vesta (Veras et al. 2017). These bodies are low mass ($M_b \lesssim 10^{21} \, \rm kg$) compared to the total accreted material inferred by accretion rates, $\sim 10^{23}\, \rm kg$ (Redfield et al. 2017). Each of these fragments has a bulk density high enough to avoid being immediately disrupted, but low enough that they undergo mantle disruption and maintain coherent structure for weeks \citep{Rappaport2017}. Further work is required to determine the likelihood of such bodies being created during early planet formation versus being formed as a product of post-main-sequence planetary system evolution \citep{2011AIPC.1331...56P, 2014A&A...563A..61S, 2018MNRAS.480.2784V}.

These suggested attributes of the asteroid's internal composition are consistent with the lack of small particles found by multiple papers \citep{Alonso2016, Zhou2016, Xu2018, Croll2017}. \citet{Xu2018} show that small grains would sublimate quickly while larger grains survive, with or without a gas-rich disk to protect them. An icy mantle with highly refractive minerals could disrupt such that the small grains sublimate to form the gas disk interior to the transiting objects while the larger grains create an extended dust cloud which produces the observed transits for long periods of time. Dynamic interactions between the debris clouds could explain the variations in transit morphology without being strong enough to disperse the dust clouds entirely. Determining when the asteroids have completely disrupted to form a largely homogeneous gas disk will place constraints on the total mass of asteroids disrupting in the system while separating the individual periodicities of the transiting objects will pin down the number of individual asteroids. If the multiple disk structure proposed by \citet{Cauley2018} continues to effectively predict the spectroscopic behavior of WD 1145+017 and we assume each individual ring is the byproduct of a single asteroid's destruction, the number and evolution of the rings \citep{Veras2015b} may provide an independent constraint on the number of disrupting bodies in addition to stability constraints imposed by simulations like those explored in \citet{Veras2016b}. Follow-up simulations and observations should try to constrain this number by fitting to blends of transit features and accounting for sublimation by determining an appropriate particle size distribution evolving over the disruption process.

The chromatic dips of Boyajian's star \citep{Boyajian2018} show some resemblance to our simulated lightcurves and may be the result of disruption without the rapid sublimation of small particles. More recent discoveries in the same vein include another transiting white dwarf reported by \citet{Vanderbosch2019} and the accretion of a gas giant envelope onto a white dwarf reported by \citet{Gansicke2019}. These are the first members of a larger class of dying planetary systems that must be studied by pairing spectroscopic and photometric observations with disruption simulations, either tidal as in WD 1145+017 or rotational as \citet{Veras2020} proposes for the body transiting ZTF J0139+5245 \citep{Vanderbosch2019}. This multi-pronged approach would use the death of these planetary systems in action to study fundamental properties of exoplanetary bodies that are otherwise inaccessible: a study in necroplanetology.

\begin{deluxetable}{ccc}
\tablecaption{The parameters common to all simulations listed in Table \ref{table:sim_param_table}.\label{table:common_params}}
\tablehead{ \colhead{Parameter} & \colhead{Value}}
\startdata
Number of Particles $(N)$ & 3000\\
Bulk Mass $(M_b)$ &$10^{24}\, \rm kg$\\
White Dwarf Mass $(M_\star)$ & $0.6 \, M_\odot$\\
Eccentricity $(e)$ & 0\\
Semimajor Axis $(a)$ & $0.0054\, \rm AU$\\
Inclination $(i)$ & 90\degree\\
\enddata
\end{deluxetable}

\begin{longrotatetable}
\begin{deluxetable*}{ccccccccc}
\tabletypesize{\scriptsize}
\tablehead{
	\colhead{Structure Number} & \colhead{$f_c$} & \colhead{$\rho_m /\rho_c$} & \colhead{Crust?} & \colhead{$f_l$} & \colhead{$\rho_l / \rho_c$} & \colhead{$\rho_b$ [$\rm g\, \rm cm^{-3}$]} & \colhead{$\mathcal{X}^2_{\nu = 110}$}\\
   \colhead{} & \colhead{Core Volume Fraction} & \colhead{Mantle/Core Density Ratio} & \colhead{} & \colhead{Crust Volume Fraction} & \colhead{Crust/Core Density Ratio} & \colhead{ Bulk Density} & {Chi-Squared with 110 degrees of freedom}}
\decimals
\tablecaption{The parameters which were varied between simulations. 10 instances of Structure 1 were run to confirm that there was no strong dependence on the random seed. 5 copies of Structure 1 with $M_b \in [10^{18}, 10^{24}]\, \rm kg$ were run to verify that the speed of disruption followed the $t_{\text{fill}}$ dependence described in \cite{Veras2017}. The chi-squared statistic of each structure's best fit is listed in the right-most column. The number of free parameters is the same for all fits so lower values of $\mathcal{X}^2_{\nu = 110}$ correspond to better fits.\label{table:sim_param_table}}
\startdata
\textbf{1\tablenotemark{a}}  &  \textbf{0.15}    & \textbf{0.25} & \textbf{No}  &  \textbf{\NA} & \textbf{\NA} & \textbf{3.0} & \textbf{126.33} \\
\textbf{2\tablenotemark{a}}  &  \textbf{0.15}    & \textbf{0.40} & \textbf{No}  &  \textbf{\NA} & \textbf{\NA} & \textbf{3.0} & \textbf{126.73} \\
3  &  0.15    & 0.55 & No  &  \NA & \NA & 3.0 & 184.73 \\
4  &  0.25    & 0.25 & No  &  \NA & \NA & 3.0 & 164.39 \\
5  &  0.25    & 0.40 & No  &  \NA & \NA & 3.0 & 167.16\\
6  &  0.25    & 0.55 & No  &  \NA & \NA & 3.0 & 189.46\\
7  &  0.35    & 0.25 & No  &  \NA & \NA & 3.0 & 167.93\\
8  &  0.35    & 0.40 & No  &  \NA & \NA & 3.0 & 161.74\\
9  &  0.35    & 0.55 & No  &  \NA & \NA & 3.0 & 179.70\\
10 &  0.15    & 0.25 & Yes &  0.1 & 0.1 & 3.0 & 146.33\\
\textbf{11\tablenotemark{a}} &  \textbf{0.15}    & \textbf{0.40} & \textbf{Yes} &  \textbf{0.1} & \textbf{0.1} & \textbf{3.0} & \textbf{128.76}\\
\textit{12\tablenotemark{b}} & \textit{ 0.15}    & \textit{0.55} & \textit{Yes} &  \textit{0.1} & \textit{0.1} & \textit{3.0} & \textit{219.39}\\
13 &  0.25    & 0.25 & Yes &  0.1 & 0.1 & 3.0 & 152.69\\
14 &  0.25    & 0.40 & Yes &  0.1 & 0.1 & 3.0 & 158.65\\
\textit{15\tablenotemark{b}} &  \textit{0.25}    & \textit{0.55} & \textit{Yes} &  \textit{0.1} & \textit{0.1} & \textit{3.0} & \textit{191.29}\\
\textit{16\tablenotemark{b}} &  \textit{0.35}    & \textit{0.25} & \textit{Yes} & \textit{ 0.1} & \textit{0.1} & \textit{3.0} & \textit{192.63}\\
\textit{17 \tablenotemark{b}}&  \textit{0.35}    &\textit{ 0.40} & \textit{Yes} &  \textit{0.1} & \textit{0.1} & \textit{3.0} & \textit{190.09}\\
18 &  0.35    & 0.55 & Yes &  0.1 & 0.1 & 3.0 & 188.69 \\ \tablebreak
\textbf{19\tablenotemark{a}}  &  \textbf{0.15}    & \textbf{0.25} & \textbf{No}  &  \textbf{\NA} & \textbf{\NA} & \textbf{4.0} &\textbf{ 131.42}\\
\textbf{20\tablenotemark{a}}  &  \textbf{0.15}    & \textbf{0.40} & \textbf{No}  &  \textbf{\NA} & \textbf{\NA} & \textbf{4.0} & \textbf{129.57}\\
21  &  0.15    & 0.55 & No  &  \NA & \NA & 4.0 & 171.49\\
22  &  0.25    & 0.25 & No  &  \NA & \NA & 4.0 & 159.53\\
23  &  0.25    & 0.40 & No  &  \NA & \NA & 4.0 & 135.95\\
24  &  0.25    & 0.55 & No  &  \NA & \NA & 4.0 & 157.27\\
25  &  0.35    & 0.25 & No  &  \NA & \NA & 4.0 & 169.05\\
26  &  0.35    & 0.40 & No  &  \NA & \NA & 4.0 & 188.30\\
27  &  0.35    & 0.55 & No  &  \NA & \NA & 4.0 & 176.13\\
28 &  0.15    & 0.25 & Yes &  0.1 & 0.1 & 4.0 & 137.11\\
29 &  0.15    & 0.40 & Yes &  0.1 & 0.1 & 4.0 & 140.48\\
30 &  0.15    & 0.55 & Yes &  0.1 & 0.1 & 4.0 & 170.63\\
31 &  0.25    & 0.25 & Yes &  0.1 & 0.1 & 4.0 & 173.21\\
32 &  0.25    & 0.40 & Yes &  0.1 & 0.1 & 4.0 & 161.53\\
\textit{33\tablenotemark{b}} &  \textit{0.25}    & \textit{0.55} & \textit{Yes} &  \textit{0.1} & \textit{0.1} & \textit{4.0} & \textit{193.29}\\
34 &  0.35    & 0.25 & Yes &  0.1 & 0.1 & 4.0 & 185.12\\
35 &  0.35    & 0.40 & Yes &  0.1 & 0.1 & 4.0 & 189.90\\
36 &  0.35    & 0.55 & Yes &  0.1 & 0.1 & 4.0 & 169.22\\
\enddata
\tablenotetext{a}{The five best-fitting structures are bold-faced. They all have a low core fraction $f_c = 0.15$ and mantle/core density ratio $\rho_m /\rho_c =0.25$ or 0.4.}
\tablenotetext{b}{The five worst-fitting structures are italicized. All five have a crust and four of them have core fractions $f_c > 0.15$. The worst-fit, Structure 12, has a low core fraction $f_c = 0.15$, but its mantle/core density ratio is high: $\rho_m /\rho_c =0.55$.}
\end{deluxetable*}
\end{longrotatetable}

\acknowledgments
The authors thank the anonymous referee for their constructive and helpful feedback. The authors also thank Wesleyan University for computer time supported by the NSF under grant number CNS-0619508 and CNS-0959856. We acknowledge support for this work from NASA Keck funds related to polluted white dwarfs associated with RSA \#1536748. G.M.D. thanks Zachory Berta-Thompson for helpful comments during the preparation of this manuscript. D.V. acknowledges the support of the STFC via an Ernest Rutherford Fellowship (grant ST/P003850/1).

%

\vspace{5mm}


\software{\texttt{astropy} \citep{2018AJ....156..123A}, \texttt{matplotlib} \citep{matplotlib}, \texttt{NumPy} \citep{numpy},  \texttt{REBOUND} \citep{REBOUND_2}, \texttt{emcee} \citep{emcee}.}

\bibliography{refs}{}
\bibliographystyle{aasjournal}
\end{document}